\documentclass{webofc}
\usepackage[varg]{txfonts}   % Web of Conferences font
\usepackage{hyperref}
\usepackage{url}
\hypersetup{colorlinks=true,citecolor=blue,urlcolor=blue,linkcolor=blue}
\usepackage{amsmath,amssymb,amsfonts}
\usepackage{algorithmic}
\usepackage{graphicx}
\usepackage{textcomp}
\usepackage{xcolor}
\usepackage{tikz}

\usepackage{listings}
\usepackage{float}

\usepackage{caption}
\usepackage{subfig} % Add this in your preamble

\usepackage{listings}
\usepackage{xcolor}

\lstdefinestyle{mystyle}{
    language=C++,
    basicstyle=\ttfamily\footnotesize,
    keywordstyle=\color{blue}\bfseries,
    commentstyle=\color{gray}\itshape,
    stringstyle=\color{red},
    numbers=left,
    numbersep=5pt,
    numberstyle=\tiny\color{gray},
    frame=single,
    breaklines=true,
    captionpos=b,
    escapeinside={(*@}{@*)}
}
\def\BibTeX{{\rm B\kern-.05em{\sc i\kern-.025em b}\kern-.08em
    T\kern-.1667em\lower.7ex\hbox{E}\kern-.125emX}}

\newcommand{\systemname}{SkimROOT}

\begin{document}

\title{\systemname: Accelerating LHC Data Filtering with Near-Storage Processing}
%
% subtitle is optional

\author{\firstname{Narangerelt} \lastname{Batsoyol}\inst{1}\thanks{\email{nbatsoyo@ucsd.edu}} \and
        \firstname{Jonathan} \lastname{Guiang}\inst{1}\and
        \firstname{Diego} \lastname{Davila}\inst{1}\and
        \firstname{Aashay} \lastname{Arora}\inst{1} \and
        \firstname{Philip} \lastname{Chang}\inst{2} \and
        \firstname{Frank} \lastname{Würthwein}\inst{2} \and
        \firstname{Steven} \lastname{Swanson}\inst{1}
        % etc.
}

\institute{University of California, San Diego, La Jolla, CA, USA
\and
           University of Florida, Gainesville, FL, USA
          }

\abstract{%
%Data analysis in high-energy physics (HEP) begins with data reduction step, where vast datasets are filtered to extract relevant physics events. At the Large Hadron Collider (LHC), this process faces significant bottlenecks  due to the physical separation of compute and storage facilities, requiring large datasets to be transferred over slow wide-area networks (WAN) for analysis. 

%To address this, we introduce \systemname, a near-data filtering system designed to accelerate LHC data analysis by leveraging Data Processing Units (DPUs) integrated into storage servers. \systemname{ } utilizes the high-bandwidth network connectivity and computational capabilities of DPUs to perform filtering operations near the data source, returning only the filtered output and significantly reducing transfer and processing delays.
%This work presents the development of \systemname, including the optimization of data filtering algorithms for a near-data computing model and insights gained from implementing this prototype.

Data analysis in high-energy physics (HEP) begins with data reduction, where vast datasets are filtered to extract relevant events. At the Large Hadron Collider (LHC), this process is bottlenecked by slow data transfers between storage and compute nodes. To address this, we introduce SkimROOT, a near-data filtering system leveraging Data Processing Units (DPUs) to accelerate LHC data analysis.  By performing filtering directly on storage servers and returning only the relevant data, SkimROOT minimizes data movement and reduces processing delays. Our prototype demonstrates significant efficiency gains, achieving a 44.3$\times $ performance improvement, paving the way for faster physics discoveries.

}
\maketitle

\section{Introduction}
\label{intro}
\vspace{-5mm}
 \begin{figure}[H]
\centering
\includegraphics[width=13cm,clip]{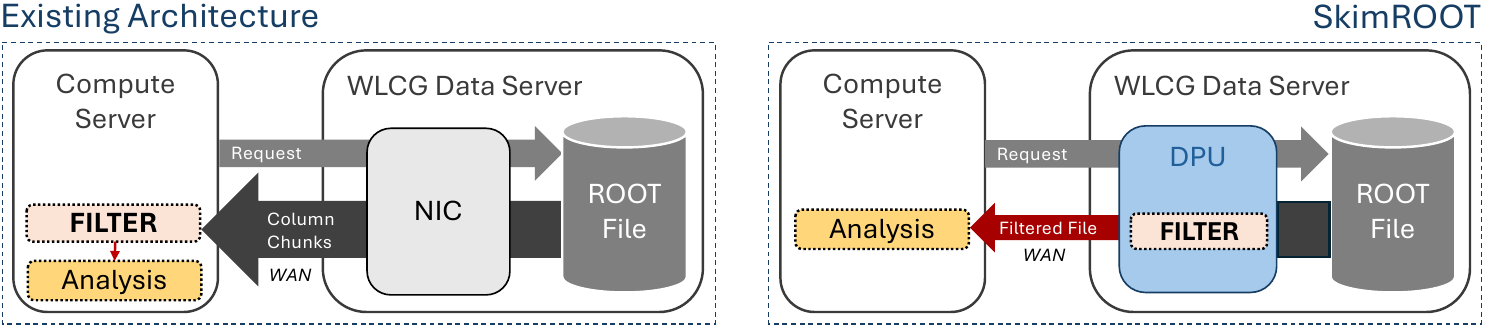}
\caption{In the legacy approach (left), the compute node fetches relevant column chunks from storage for filtering, resulting in high data movement.  With \systemname{} (right), the DPU filters data within the storage server, returning only the filtered data.}
\label{fig:intro}       % Give a unique label
\end{figure}
\vspace{-5mm}

%The Large Hadron Collider (LHC) at CERN has been essential in advancing our knowledge of particle physics, enabling exploration of the universe’s fundamental structure~\cite{cern2024}. 

The Large Hadron Collider (LHC) at CERN advances our knowledge of particle physics by colliding high-energy protons, generating subatomic particles recorded by detectors like CMS~\cite{cms2008cms} and ATLAS~\cite{aad2008atlas}. %These experiments explore fundamental forces of the universe and search for new physics. 
Each year, the LHC generates over 100~PB~\cite{CERNDataCentre2021} of particle collision data, managed and distributed through the Worldwide LHC Computing Grid (WLCG), where compute and storage resources are spread across multiple globally connected sites.

To analyze these vast datasets, scientists first perform data filtering (or skimming)~\cite{gutsche2018cms, graur2023evaluating}, reducing dataset size by extracting only the relevant collision data for specific studies. Scientists submit filtering jobs to WLCG, where a workload management system schedules jobs on available computing clusters, which may be at different sites from where the data is stored. When filtering jobs request data stored at a remote site, large transfers over WANs introduce delays and increase network overhead. 

Moreover, job management in WLCG is complex—jobs frequently fail and require resubmission. For CMS~\cite{cms2008cms}, skimming can take days to weeks for a single analysis~\cite{gutsche2018cms, galewsky2020servicex}, depending on data location and resource availability. In the HL-LHC~\cite{bruning2024high}, where data volumes are projected to increase nearly tenfold~\cite{CMSComputingReport2022}, accelerating filtering is even more important.

LHC data is stored in a complex, compressed columnar format, where filtering extracts specific collision data based on user-defined criteria. However, remote data access significantly impacts filtering performance, especially when filtering code is inefficient, resulting in excessive data transfers~\cite{xroot}. Writing efficient filtering queries, especially in C++, is challenging and requires significant expertise in low-level optimization.

In this paper, we introduce \systemname{}, a near-storage processing system that accelerates LHC data filtering and reduces analysis turnaround times, enabling faster physics discoveries. By leveraging Data Processing Units (DPUs), \systemname{} filters data directly at the storage source before transmission (Figure 1, right), significantly reducing data movement and network overhead. A DPU is a specialized add-in network interface card (NIC) designed for both network data transfer and on-device processing. It connects to the host server via PCIe and integrates high-speed networking, power-efficient CPUs, and dedicated accelerators for networking and computation, enabling it to offload tasks from the host processor.

% By offloading these operations from the host processor, DPUs improve efficiency and reduce system bottlenecks.

%DPUs are a class of SmartNICs that connect to host servers via PCIe and can function as standalone servers. They integrate high-speed networking, power-efficient CPUs, and dedicated accelerators for both networking and computation.

This work presents the first prototype of LHC data filtering on DPUs, demonstrating their feasibility for near-storage processing. To simplify user queries, SkimROOT introduces a JSON-based query format, allowing users to define selection criteria without complex C++ scripting. Filtering is executed using a two-phase model that minimizes I/O by dynamically loading only required branches. The DPU processes the request directly at the storage source, applies the filtering conditions, and returns a reduced dataset. Additionally, we provide the first detailed breakdown of LHC data filtering performance, analyzing decompression overhead and I/O latency to identify key bottlenecks in traditional workflows.
%This work presents the first prototype of LHC data filtering on DPUs, demonstrating their feasibility for near-storage processing. We introduce an optimized two-step filtering model that dynamically loads only the necessary branches at each stage, reducing unnecessary data transfers. To make filtering more accessible, \systemname{} allows users to define selection criteria in a simple JSON format and submit filtering requests to the DPU. 

Our evaluation demonstrates that SkimROOT accelerates filtering by 44.3× compared to client-side filtering and reduces data fetch time by 3.18× over server-side filtering on LZ4-compressed files. The remainder of this paper is structured as follows: Section~\ref{sec:background} discusses LHC data filtering and DPUs. Section~\ref{sec:design} outlines \systemname{}'s design and implementation. Section~\ref{sec:evaluation} evaluates \systemname{}'s performance, and Section~\ref{sec:conclusion} concludes the paper.

%\systemname{} configures the DPU as an independent server within the storage node, executing filtering logic while leveraging its hardware decompression engine for efficient basket decompression. This reduces CPU overhead on compute nodes and minimizes I/O delays.

%To further streamline analysis, \systemname{} provides an HTTP POST endpoint, allowing scientists to submit filtering requests without complex C++ scripting. Additionally, \systemname{} optimizes filtering code and ROOT data access, ensuring that only necessary data is processed and transferred.

%Our work makes the following key contributions, along with identifying filtering bottlenecks through the first detailed breakdown of LHC data filtering performance, analyzing decompression overhead and I/O latency.
%\begin{itemize}
%\item \textbf{Prototyping DPUs for LHC Data Filtering:} We demonstrate the first implementation of LHC data filtering on DPUs integrated into storage systems.
%\item \textbf{Optimized Filtering:} We introduce a two-step filtering model that minimizes unnecessary data transfers by dynamically loading only the required branches at each processing stage.
%\item \textbf{Ease of use:}   \systemname{} provides a REST API, allowing users to perform efficient filtering without modifying storage server APIs or writing complex C++ scripts.
%\end{itemize}

\vspace{-3mm}
\section{Background}
\label{sec:background}
\systemname{} accelerates LHC data filtering by leveraging a DPU for near-data processing. This section outlines the LHC data layout, filtering process and relevant DPU background.

%Filtering selects events by applying user-defined criteria to specific columns. ROOT retrieves only the baskets for selected columns, avoiding full file loads (Figure \ref{fig:intro}, left). However, performance degrades when jobs access data from remote storage, as each required basket must be transferred individually. Although ROOT employs prefetching mechanisms to aggregate small requests, bulk transfers over long distances still introduce delays.
%Additionally, inefficient filtering logic leads to excessive data transfers, increasing I/O overhead~\cite{xroot}.

\vspace{-2mm}
\subsection{ROOT file and I/O} \label{sec:io}

\begin{figure}[h]
   \centering
   \subfloat[][\texttt{TTree} structure]
    {\includegraphics[width=0.29\linewidth]{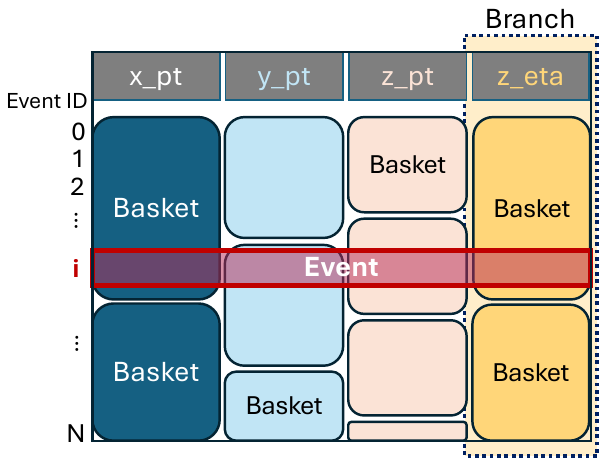} \label{fig:root}}
    \hfill
    \subfloat[][Original Query]
    {\includegraphics[width=0.33\linewidth]{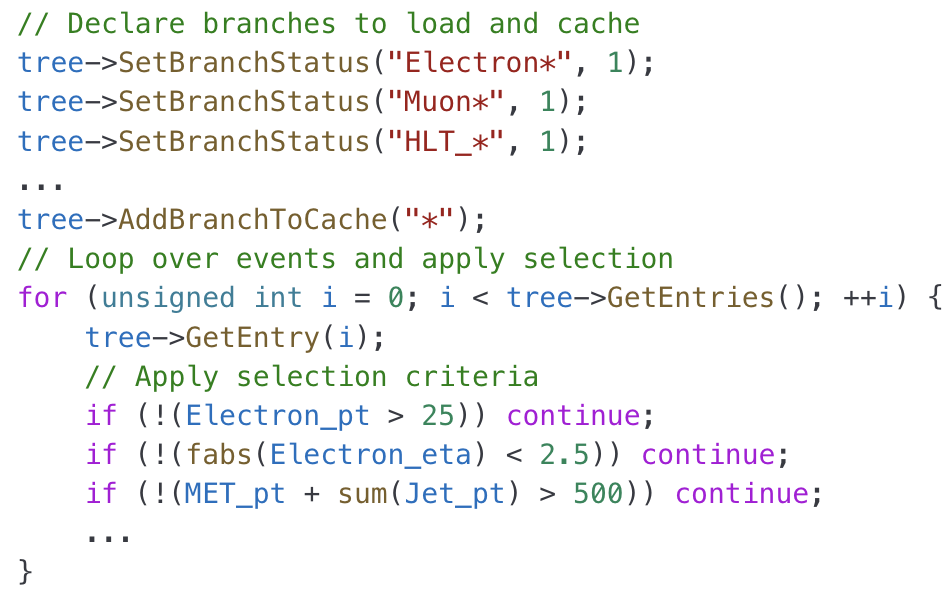} \label{fig:query}}
    \hfill
    \subfloat[][SkimROOT's Query]
   {\raisebox{7mm}{\includegraphics[width=0.31\linewidth]{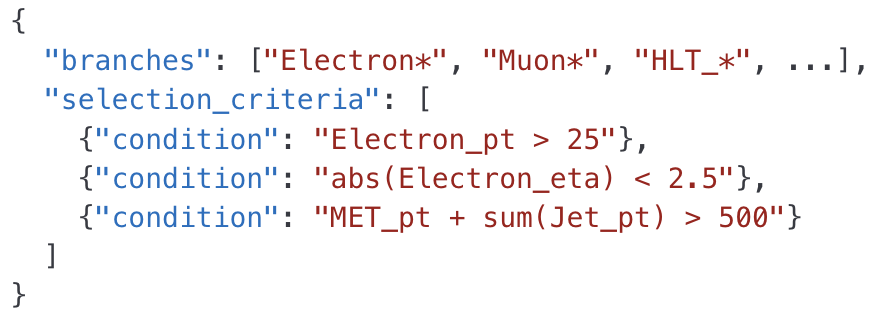}} \label{fig:json}} % Moves it up by 5mm
\vspace{-1mm}
   \caption{(a) TTree, (b) Original Filtering and (c) SkimROOT's JSON-based User Query.}
    \label{fig:filtering_comparison}
    \vspace{-8mm}
\end{figure}

LHC data is processed using ROOT, the most widely used framework in HEP~\cite{ROOTPaper}. ROOT represents the physics properties of collision-produced particles as C++ objects and stores them in ROOT files. These files use a columnar format with \texttt{TTree} (see Fig.~\ref{fig:root}), where each row corresponds to a collision event, and columns ("branches") store particle properties such as momentum, energy, and charge. To optimize I/O, ROOT compresses consecutive column entries into "baskets," the fundamental unit for data access and compression.

ROOT files vary in format depending on the required level of detail. Since Run 3, the NanoAOD format has been the most widely used for CMS analyses. It typically ranges from 1–4 GB and contains 1–2 million events with 1000–2000 branches.

NanoAOD branch names follow a structured naming convention, grouping related variables under common prefixes (e.g., \texttt{Electron\_pt}, \texttt{Electron\_eta}). This allows users to select entire groups of branches using wildcards instead of explicitly specifying each branch.

 % To optimize I/O, ROOT organizes consecutive entries from the same branch into compressed chunks called baskets (typically 32 KB before compression). A basket, represented by the \texttt{TBasket} class, serves as the basic unit for I/O and compression. 

%Each \texttt{TTree} consists of multiple branches (\texttt{TBranch} objects), each referencing an array of \texttt{TBasket} objects that store event data in binary format. To enable efficient data retrieval, each \texttt{TBranch} maintains an array mapping baskets to their first event ID, allowing quick lookup of the basket containing a given event. Within each \texttt{TBasket}, an offset array indexes event positions, enabling direct access to specific event data without scanning the entire basket.

\vspace{2mm}

\noindent\textbf{Reading data in ROOT.} When reading event data, ROOT relies on metadata stored in the file header, including object descriptions, column offset indices, and type information necessary for interpreting and deserializing stored data. To retrieve the \texttt{i}-th event from a branch, ROOT locates the corresponding branch and determines which basket contains the event by referencing its ``first event index array", which lists the starting event ID for each basket in the branch.
The basket is then loaded into memory and decompressed. Using the event offset array within the basket, ROOT directly accesses the event’s binary data and deserializes it into a C++ object using metadata from the file header.

%When reading event data, ROOT relies on metadata stored in the file header, including object descriptions, an offset index to columns, and type information needed to interpret and deserialize stored data. To retrieve the \texttt{i}-th event from a branch, ROOT follows these steps:
%\begin{enumerate}
%\item \textit{Identify the Branch and Locate the Basket:} Find the requested branch  and use its first event index array to determine which basket contains the \texttt{i}-th event.
%\item  \textit{Load and Decompress:} Load the compressed basket into memory and decompress it.
%\item \textit{Locate and Deserialize:} Use the basket's event offset array to find the \texttt{i}-th event's binary data and decode it into a C++ object using metadata from the file header.
%\end{enumerate}

%At the LHC, high-energy proton-proton collisions produce subatomic particles, recorded by detectors like CMS and ATLAS. These experiments analyze particle trajectories, momenta, and energy to study fundamental forces and search for new physics.

\vspace{-3mm}
\subsection{LHC Data Access and Filtering}
\vspace{-2mm}
Access to storage clusters at WLCG is managed by frameworks like XRootD, which, along with its protocol, facilitates efficient data retrieval. Compute nodes running filtering jobs request data via the XRootD client, which forwards the request to an XRootD server, typically hosted on a data transfer node (DTN) within the storage cluster. The DTN retrieves the requested data from storage backends such as a disk pool or a distributed file system like EOS or Ceph, and the XRootD server streams it back to the compute node for processing.

Scientists filter ROOT files by writing C++ or Python scripts that extract relevant events, significantly reducing dataset size—often by orders of magnitude. ROOT retrieves only the baskets for selected columns, avoiding full file loads (Figure \ref{fig:intro}, left). However, performance degrades when jobs access data from remote storage, as each required basket must be transferred individually. Although ROOT employs prefetching through TTreeCache to aggregate small requests over the network, bulk transfers over long distances still introduce delays.

The filtering process begins by opening a ROOT file and reading its header. Users specify the required branches, often using wildcard expressions for simplicity. As shown in Fig.~\ref{fig:query}, the filtering script iterates over events, reading only the specified branches to reconstruct the \texttt{TTree}. For each event, \texttt{tree->GetEntry(i)} retrieves, decompresses, and extracts the relevant baskets, populating the declared branches for the \texttt{i}-th row. The script then applies user-defined selection criteria and writes only the filtered events to an output ROOT file.

Since filtering processes events sequentially and ROOT stores data in a columnar format, retrieving a single event often requires accessing multiple non-contiguous baskets. While \texttt{TTreeCache} prefetches baskets for selected branches over a range of entries, its efficiency depends on data locality and access patterns and may not always provide optimal results.

%Access to storage clusters at LHC data centers is enabled by frameworks like XRootD, which, along with the XRootD protocol, facilitates efficient data retrieval for filtering and analysis. Compute nodes running filtering jobs request data via the XRootD client, which forwards the request to an XRootD server, typically hosted on a data transfer node (DTN) within the storage cluster. The DTN retrieves the requested data from the storage backend, such as a disk pool or a distributed system like EOS or Ceph, and the XRootD server streams the data back to the compute node for processing.

\subsection{Data Processing Unit (DPU)}
\label{dpu}
A DPU is an advanced SmartNIC designed to offload networking and security functions in data centers, reducing the computational burden on host CPUs. It operates independently with its own operating system, memory, power-efficient processors, and specialized hardware accelerators for tasks such as pattern matching, decompression and encryption.

The NVIDIA BlueField-3 (BF-3)~\cite{nvidia_bluefield3_datasheet}, used in our prototype, runs Ubuntu and is equipped with 16 ARM Cortex-A78 processors, 32 GB of DRAM, and 128 GB storage. It integrates a ConnectX-7 NIC with 400 Gb/s external networking. Additionally, it supports PCIe Gen 5.0 x32, delivering up to 1 Tbps of bandwidth to the host. The DPU supports SSH access from the host and can run applications natively or within Docker containers.

 BF-3's decompression engine supports DEFLATE and LZ4 algorithms. The DOCA~\cite{nvidia_doca_sdk} framework enables programming and offloading tasks to its acceleration engines.

DPUs can operate in two modes. In ``Embedded Mode", the DPU functions as a packet-processing intermediary, handling all ingress and egress traffic with the ability to perform packet-level modifications. In ``Separated Host Mode", it operates as a standalone server with its own IP and MAC address, allowing traffic to bypass the DPU when reaching the host while enabling the execution of general-purpose compute applications.

\section{Design \& Implementation}
\label{sec:design}
%\systemname{} configures the DPU as an independent server within the storage node, executing filtering logic while leveraging its hardware decompression engine for efficient basket decompression. 

\systemname{} accelerates LHC data skimming by filtering data near storage, eliminating the need to transfer and process data on WLCG compute nodes over a high-latency network.
By leveraging near-storage processing with a DPU, \systemname{} offloads filtering from compute nodes, freeing resources for other tasks and accelerating scientific discoveries.

 \begin{figure}[H]
\centering
\includegraphics[width=13cm,clip]{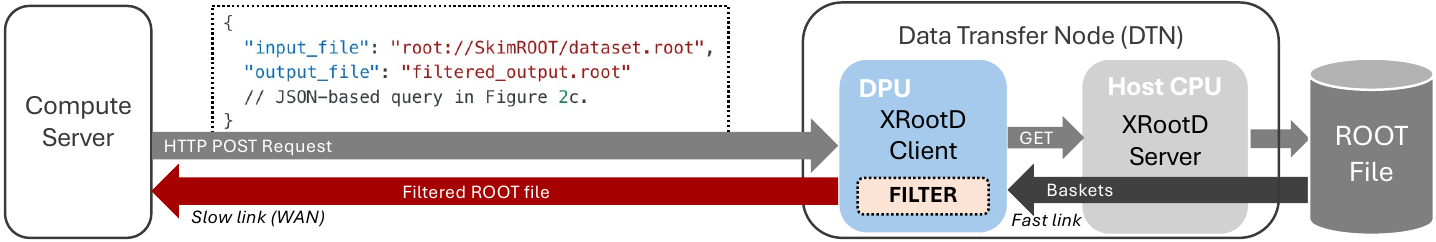}
\caption{\systemname{} Overview.}
\label{fig:overview_skimroot}   
\vspace{-3mm}% Give a unique label
\end{figure}

In a typical \systemname{} workflow, as shown in Figure \ref{fig:overview_skimroot}, the XRootD server running on the data transfer node (DTN) within the storage cluster provides access to large ROOT files stored in the backend storage. Instead of transferring data to compute nodes for filtering, the DPU—connected to the DTN via PCIe—acts as an XRootD client, running the filtering program and returning only the reduced final ROOT file to the compute node.

To simplify user queries, SkimROOT replaces traditional ROOT-based filtering scripts with a structured JSON query format (Figure \ref{fig:json}). Instead of writing manual C++ scripts, users define selection criteria in JSON, eliminating the need for ROOT-specific programming expertise while ensuring a flexible and human-readable event selection process.

Configured in "Separated Host" mode, the DPU operates independently with its own IP address, enabling it to receive HTTP POST requests with user-defined selection criteria. Upon receiving a request, the filtering program running on the DPU's ARM cores parses the JSON payload and retrieves the necessary baskets from the XRootD server, and executes the filtering. To accelerate decompression, \systemname{} leverages the DPU’s hardware decompression engine, reducing CPU overhead traditionally associated with reading baskets.

\systemname{} optimizes filtering execution by reducing unnecessary data transfers. Traditional filtering retrieves excessive branches due to broad wildcard selections and inefficient filtering logic. While wildcard selection simplifies scripting, it often loads more data than needed. \systemname{} dynamically enables only the required branches for filtering and output, minimizing data movement and memory usage. Additionally, it optimizes branch access by loading only essential variables at each filtering stage, ensuring efficient data retrieval.

\subsection{User Query and Branch Selection Optimization}

To interact with \systemname{}, users submit filtering requests via HTTP POST, specifying the input dataset, output file, selection criteria, and required branches in a structured JSON payload. These requests can be made using \texttt{curl}, with selection criteria provided either directly in the command or through an external JSON file. This request-driven approach enables users to define event selection conditions at a high level without handling low-level ROOT operations. Upon receiving a request, \systemname{} parses the explicitly declared branches and selection criteria, including numerical thresholds, logical conditions, and dependencies, before dynamically determining data access and processing strategies.

A major inefficiency in traditional filtering is the retrieval of unnecessary data. Typically, all selected branches are loaded for every event, even though only a subset is required for filtering. In NanoAOD-based analyses, O(10) branches are used for skimming, while O(100) are included in the final output. Instead of loading all branches upfront, \systemname{} categorizes them into two groups: filtering criteria branches, used to determine event selection, and output-only branches, retrieved only if an event passes the selection criteria.

Traditional filtering also increases data movement due to broad wildcard selections. Users often specify patterns like HLT\_* to simplify scripting and include all potentially relevant branches.%%However, CMS analysis shows that while HLT\_* expands to over 650 branches, most physics analyses rely on fewer than 23 specific triggers. 
However, based on common practice in CMS analyses, while HLT\_* expands to over 650 branches, most physics studies typically rely on fewer than 23 specific triggers. To prevent excessive retrieval, \systemname{} maps wildcard selections to a minimal, predefined branch set based on usage statistics. If users require all branches, they can override this behavior by setting "force\_all": true". If the predefined branch set is used, SkimROOT logs a warning for any missing branches that were excluded due to optimization. By dynamically refining selections, \systemname{} prevents unnecessary data transfer while ensuring that relevant branches are retained.

\subsection{Optimized Filtering Execution}

After parsing the user query and dynamically enabling necessary branches, \systemname{} executes the filtering program on the DPU's ARM cores, leverages its high-bandwidth connection to the server to fetch relevant baskets while offloading decompression to its accelerator. 

\systemname{} follows a two-phase execution model that optimizes I/O efficiency by deferring non-essential data retrieval. In the first phase, it loads only filtering criteria branches, evaluating events without retrieving unnecessary data. Events that fail the selection conditions are discarded immediately. If an event passes, the second phase retrieves output-only branches only after selection is complete before writing the event to the final ROOT file.

% \systemname{} processes events using a two-phase execution model: first processing filtering criteria branches and, only after an event is selected, retrieving output-only branches. 
%The DPU executes the ROOT \texttt{branch->GetEntry(i)} function for each required branch on the i-th event, fetching only the minimal subset of data needed to evaluate selection criteria.
%Each event is then evaluated against the user-defined filtering conditions. Events failing the selection criteria are immediately discarded. If an event passes, \systemname{} retrieves the additional output-only branches and writes the event to the final ROOT file. By dynamically enabling only the required branches and deferring non-essential data until after filtering, \systemname{} optimizes memory usage and minimizes I/O overhead.

\systemname{} applies a structured, multi-step filtering model that progressively discards events at different stages. This hierarchical approach adapts to diverse selection criteria, providing flexibility across different physics analyses. Filtering starts with preselection, discarding events that do not meet basic criteria, such as requiring at least one high-quality lepton. These checks involve evaluating a single branch with simple operator conditions, ensuring minimal overhead. Events passing this stage proceed to object-level selection, where individual particles—such as electrons, muons and jets—are evaluated based on user-defined kinematic and identification criteria, which require multi-column data processing. At the event level, selection is further refined using composite variables such as the scalar sum of transverse momenta, trigger conditions, or other derived more complex calculations. This structured execution model efficiently eliminates irrelevant events early, deferring the loading of non-filtering branches until final selection is complete.

\vspace{-3mm}
\section{Evaluation}
\label{sec:evaluation}
We evaluate SkimROOT’s performance in a filtering task required for a real-world Higgs physics analysis conducted at UCSD, comparing it to client-side and server-side filtering. Our analysis examines how much SkimROOT improves filtering efficiency and what drives its performance gains. By measuring end-to-end latency, data transfer efficiency, and compute resource usage, we demonstrate the benefits of near-storage processing with DPUs.

The setup includes an XRootD server (Intel Xeon Gold 6230) hosting ROOT files, a client node submitting HTTP POST requests, and a BF-3 DPU executing filtering. The server and DPU communicate over a 128 Gb/s PCIe link, limited by the server’s PCIe Gen 3.0. 

The evaluation uses a NanoAOD file with 1749 branches, where 27 branches are used for filtering and 89 are required in the final output. A 100 MB TTreeCache is used in all methods to optimize data retrieval. To isolate the efficiency of different filtering approaches, performance is measured using a single-threaded job on a single core.

%Performance is assessed under two network conditions: a 400 Mbps setting, simulating remote file fetch speeds at Tier-2 facilities, and a 10 Gbps high-speed connection for local storage access. At UC San Diego’s Tier-2 facility, worker nodes share a 10 Gbps link, with per-job bandwidth averaging 100 Mbps, making network speed a key bottleneck. \textit{Wondershaper} is used to throttle network interface bandwidth for controlled evaluation.

We evaluate performance under three network conditions: 1~Gbps represents realistic remote WAN bandwidth on dedicated research networks, 10~Gbps models shared storage access at UCSD’s Tier-2 facility, and 100~Gbps corresponds to high-performance dedicated storage access, typically available at Tier-1 centers. The 1~Gbps case is the primary focus, as it best reflects real-world constraints for remote data access in WLCG computing. We use Wondershaper~\cite{wondershaper} to throttle network bandwidth for controlled evaluation.

%To assess performance under realistic network conditions, two scenarios are considered. A slow network setting, simulating shared bandwidth at a Tier-2 facility, is configured at 400 Mbps, which reflects typical remote file fetch speeds. A high-speed 10 Gbps connection represents local storage access in a high-performance computing environment. At UC San Diego’s Tier-2 facility, worker nodes share a 10 Gbps link to storage, with an observed per-job bandwidth of approximately 100 Mbps, making network bandwidth a key bottleneck. Wondershaper is used to throttle network bandwidth accordingly for controlled performance evaluation.

\vspace{2mm}
\noindent\textbf{Latency Analysis. }
We measure end-to-end latency from request submission to receiving the filtered file using a NanoAOD file, compressed to 3GB with LZMA and 5GB with LZ4. The evaluation compares client-side filtering with LZMA and LZ4, an optimized filtering on client-side, and SkimROOT, where the DPU processes and returns the filtered ROOT file. 

\begin{figure}[h]
    \centering
    \subfloat[][Filtering latency across different network speeds, with \systemname{} maintaining consistently low latency.]
    {\includegraphics[width=0.52\linewidth]{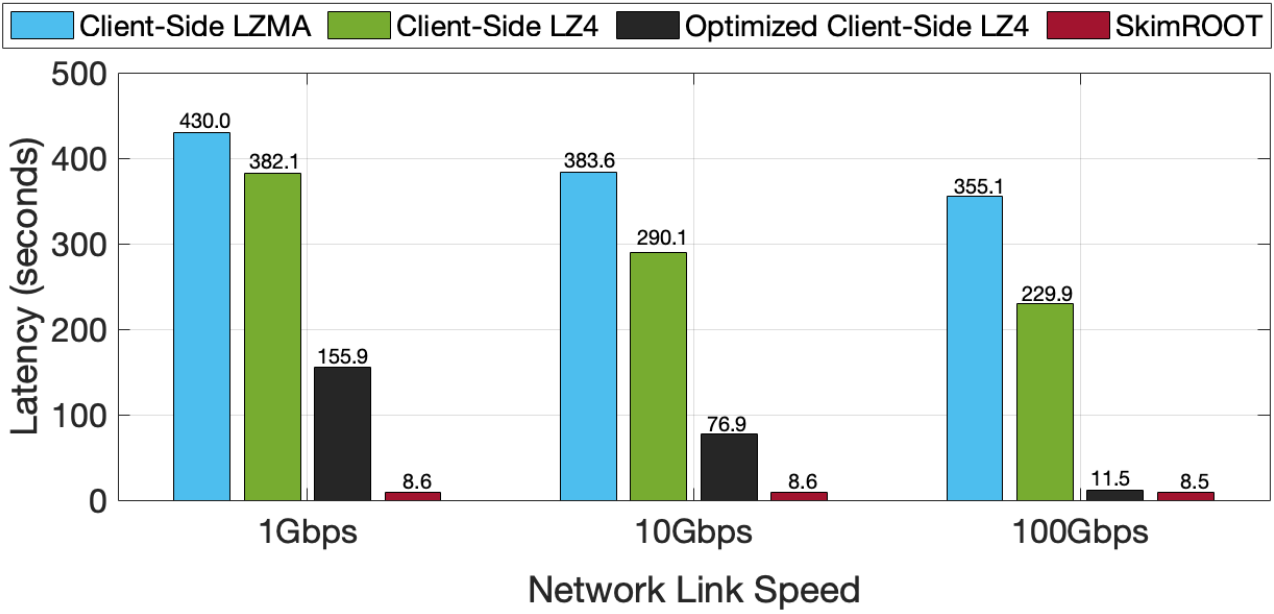} \label{fig:plot1}}
    \hfill
    \subfloat[][Breakdown of filtering latency for different methods over a 1~Gbps client-server link.]
    {\includegraphics[width=0.465\linewidth]{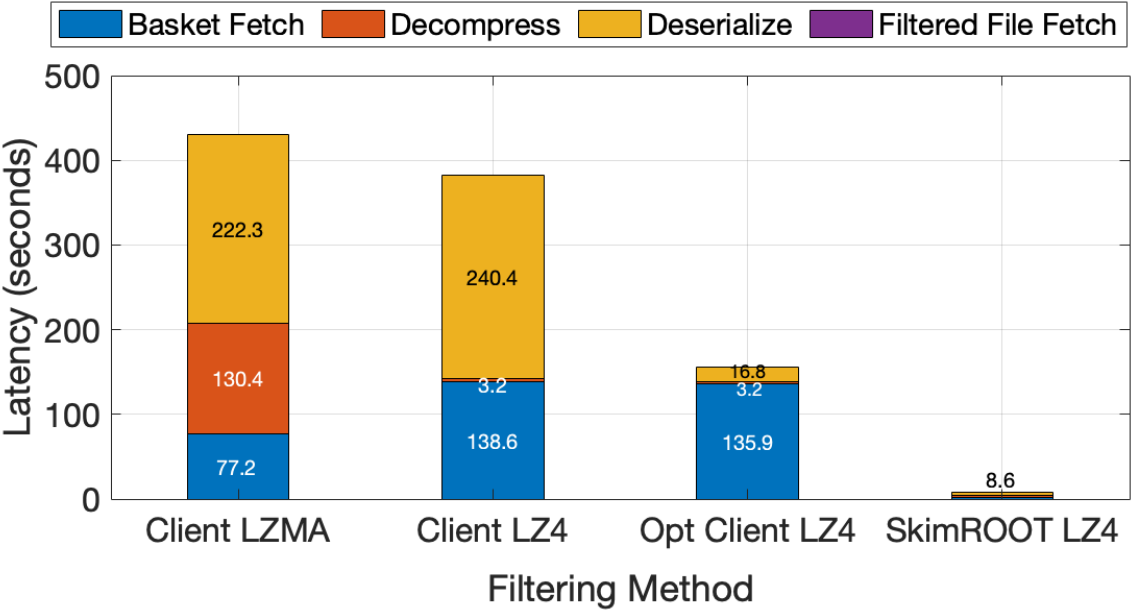} \label{fig:plot2}}
\vspace{-2mm}
   \caption{Filtering performance: (a) Overall latency. (b) Execution time breakdown.}
    \label{fig:filtering_comparison}
    \vspace{-4mm}
\end{figure}

Figure \ref{fig:plot1} presents filtering performance across different network speeds. At 1~Gbps, client-side filtering with the LZMA-compressed file takes 430s, while using the LZ4-compressed version reduces this to 382.1s due to faster decompression. Implementing our two-phase execution model in “Client Opt LZ4”  further reduces latency to 155.9s. Moving the filtering to the DPU with SkimROOT achieves the best performance at just 8.62s, delivering a 44.3× speedup over unoptimized client-side filtering with LZ4.

As bandwidth increases, client-side filtering performance improves. For example, at 100~Gbps, “Client Opt LZ4”  completes in 11.5s. However, even at higher bandwidths, SkimROOT continues to outperform client-side filtering, benefiting from high-bandwidth storage access and hardware-accelerated decompression.

\noindent\textbf{Operation Breakdown.} Figure~\ref{fig:plot2} presents a breakdown of execution time by operation over a 1~Gbps client-server link, highlighting inefficiencies in client-side filtering. Basket fetch, decompression, and deserialization collectively dominate the execution time, while SkimROOT introduces an additional step for transferring the filtered output to the client, which is negligible due to the small output size.

Although LZMA minimizes transfer size, it incurs a substantial decompression overhead of 130.4s. LZ4 alleviates this bottleneck, completing decompression in just 3.2s, but still incurs a long deserialization time (240.4s) due to inefficient filtering logic that loads unnecessary branches. While LZ4’s faster decompression improves compute efficiency, its benefit is partially offset by the larger transfer size for the same baskets, resulting in more data to deserialize and transmit. “Client Opt LZ4”  mitigates this by reducing deserialization time to 16.8s,  yet basket fetch remains a major bottleneck (135.9s) due to inefficient \texttt{TTreeCache} prefetching of non-contiguous data for randomly accessed output-only branches over the 1~Gbps link.

%\noindent\textbf{Operation Breakdown. }Figure \ref{fig:plot2} breaks down execution time by operation over a 1~Gbps client-server link, exposing inefficiencies in client-side methods. Basket fetch, decompression, and deserialization dominate execution time, while SkimROOT introduces an additional step of sending the filtered output to the client. 

%While LZMA reduces transfer size, it incurs a 130.4s decompression overhead. LZ4 removes this bottleneck, decompressing in just 3.2s, but still suffers from high deserialization costs (240.4s) due to inefficient filtering logic which fetches unnecessary data. Although LZ4's decompression is significantly faster, its advantage in remote file fetching is diminished by a larger transfer size for the same baskets. ``Client Opt LZ4" mitigates this by reducing deserialization time to 16.8s,  yet basket fetch remains a bottleneck due to network link speed and inefficient \texttt{TTreeCache} prefetching non-contiguous data for randomly accessed output-only branches.

%SkimROOT eliminates these inefficiencies across all stages and Figure \ref{fig:plot3} provides better visibility: basket fetch time drops from 139s to just 2.3s when running filtering on DPU, decompression is reduced to 2.2s with the DPU's hardware accelerator, and deserialization remains at 4.1s as we are loading same amount of data with ``Client Opt LZ4". Filtered file fetch adds only 0.02s, as the output file is only 5.2MB.

\begin{figure}[h]
    \centering
    \subfloat[][]
    {\includegraphics[width=0.55\linewidth]{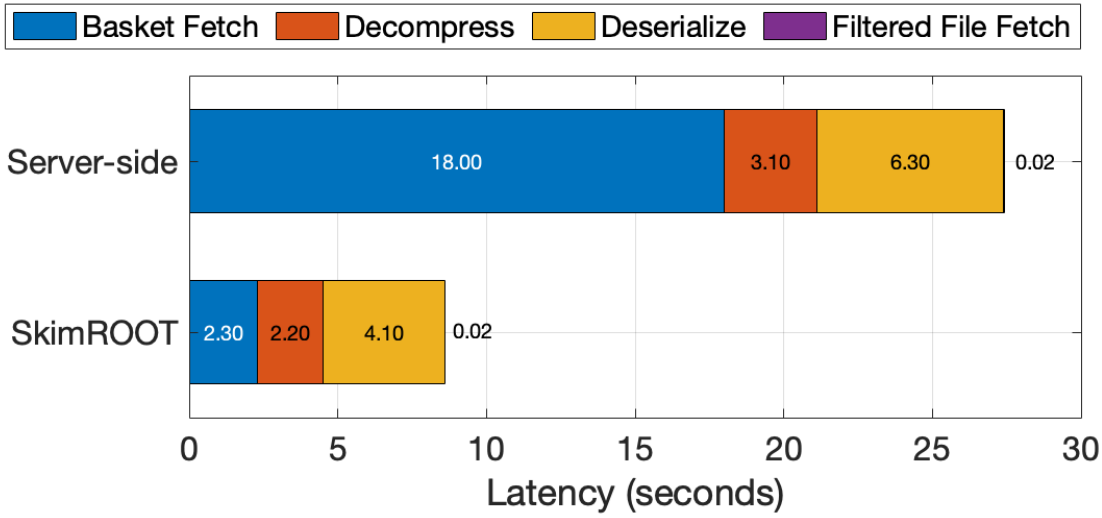} \label{fig:plot3}}
    \hfill
    \subfloat[][]
    {\includegraphics[width=0.42\linewidth]{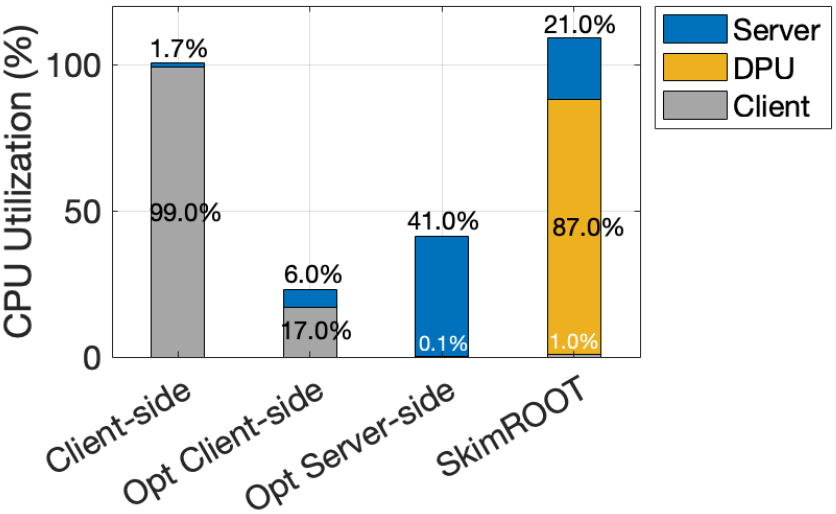} \label{fig:plot4}}
\vspace{-3mm}
   \caption{ (a) Execution time breakdown. (b) CPU Utilization (\%) for different methods.}
    \label{fig:filtering}
    \vspace{-5mm}
\end{figure}

\vspace{2mm}
\noindent\textbf{Near-Storage Filtering Latency. }To evaluate the impact of near-storage filtering, we compare SkimROOT with server-side optimized filtering in Figure \ref{fig:plot3}, where filtering is performed directly on the XRootD server. While this eliminates network transfer overhead by reading baskets from local storage instead of fetching them over XRootD, server-side filtering still incurs an 18s basket fetch time, compared to just 2.3s with \systemname.

\vspace{-2mm}

%The primary limitation is that \texttt{TTreeCache} does not function for local ROOT file access, preventing ROOT from prefetching and batching reads as it does with remote XRootD access. As a result, baskets are loaded individually, causing frequent random disk accesses and higher I/O overhead. In contrast, SkimROOT leverages XRootD’s prefetching to reduce basket fetch latency and accelerates decompression from 3.1s to 2.2s using the DPU’s hardware engine.

%While both methods process the same filtered output, server-side filtering shows a slightly higher deserialization time (6.3s vs. 4.1s), due to the lack of \texttt{TTreeCache} and synchronous basket access. ROOT reads and decompresses each basket on demand, causing the deserialization loop to pause between baskets. In contrast, SkimROOT benefits from XRootD prefetching and hardware-accelerated decompression, maintaining a steady data flow and reducing overall latency.

A key limitation of server-side filtering is that \texttt{TTreeCache} does not function for local ROOT file access, preventing ROOT from prefetching and batching reads as it does with remote XRootD. As a result, baskets are read and decompressed on demand, one at a time, leading to frequent random disk accesses and increased I/O overhead. This also affects deserialization, which must wait for each basket to become available before processing can continue, resulting in a higher deserialization time (6.3s vs. 4.1s) despite identical filtered output. In contrast, SkimROOT leverages XRootD prefetching and accelerates decompression from 3.1s to 2.2s using the DPU’s hardware engine, keeping the deserialization loop continuously supplied with data and reducing overall latency.

%This difference is attributed to increased wait time in the processing pipeline due to delayed or unbuffered basket delivery. 
%By shifting filtering from the XRootD server to the DPU, SkimROOT overcomes local storage inefficiencies, leverages XRootD prefetching, reduces I/O overhead, and achieves significantly faster filtering times.

Finally, the filtered file fetch time is negligible (0.02s) in both cases due to the small 5.2~MB output. By shifting filtering from the XRootD server to the DPU, SkimROOT overcomes local storage inefficiencies, minimizes pipeline stalls, and leverages caching and hardware acceleration to achieve significantly lower end-to-end latency.

%Finally, filtered file fetch time remains negligible (0.02s) for both configurations due to the small output size (~5.2 MB).
%By shifting filtering from the XRootD server to the DPU, SkimROOT overcomes local storage inefficiencies, minimizes pipeline stalls, and leverages caching and decompression hardware to achieve significantly faster filtering performance.

%The primary limitation is that \texttt{TTreeCache} does not function for local ROOT file access, preventing ROOT from prefetching and batching reads as it does with remote XRootD access. Instead, baskets are loaded individually, leading to frequent random disk accesses and increased I/O overhead. In contrast, SkimROOT optimizes access patterns while leveraging XRootD’s caching mechanisms, significantly reducing basket fetch latency.

%Furthermore, SkimROOT accelerates decompression from 3.1s to 2.2s using the DPU’s hardware engine, while deserialization remains at 4.1s for both methods, as they process the same filtered output. The filtered file fetch time is negligible (0.02s) in both cases due to the small 5.2 MB output.
%By shifting filtering from the XRootD server to the DPU, SkimROOT overcomes local storage inefficiencies, leverages XRootD prefetching, reduces I/O overhead, and achieves significantly faster filtering times.

\vspace{2mm}
\noindent\textbf{Resource Utilization.}
Figure~\ref{fig:plot4} shows CPU utilization per core for different filtering methods using the LZ4-compressed file over a 1~Gbps link. Original client-side filtering saturates the client CPU at 99\% and takes 382.1s to complete. Although optimized client-side filtering reduces CPU usage to 17\% by avoiding unnecessary deserialization, it still requires 155.9s—18× longer than \systemname{}—due to high basket fetch latency (135.9s vs. 2.3s). Server-side filtering eliminates network transfer overhead and reduces client CPU usage to just 0.1\%, but increases server utilization to 41\% and remains 3.18× slower than \systemname.

SkimROOT minimizes client and server CPU usage by offloading filtering to the DPU, which operates at 87\%, while the XRootD server remains at 21\%. Our evaluation shows that BF-3’s ARM cores perform comparably to host CPUs while being more power efficient. By offloading filtering, SkimROOT reduces CPU overhead on both the client and XRootD server, achieving faster processing while minimizing compute and network bottlenecks.

\vspace{-2mm}

\vspace{-2mm}
\section{Conclusion}
\label{sec:conclusion}
SkimROOT introduces a near-storage processing paradigm for LHC data filtering, leveraging DPUs to significantly accelerate event selection while reducing compute and network overhead. Our evaluation demonstrates that SkimROOT achieves up to a 44.3× speedup over original client-side filtering on a 1~Gbps link and is 3.18× faster than server-side filtering on LZ4-compressed files. These performance gains result from optimized filtering logic, hardware-accelerated decompression, and high-bandwidth storage access.

Future work will explore advanced data prefetching strategies, improved parallelization, and scalability across multiple DPUs to further enhance SkimROOT's efficiency. By integrating near-storage processing into LHC computing workflows, SkimROOT serves as a scalable and efficient prototype for managing the rapidly increasing data volumes expected in the HL-LHC era and beyond.

\vspace{-2mm}

\bibliography{ref}
%\begin{thebibliography}{}
 % Use the appropriate .bst file
% Assumes your file is named ref.bib

%Journal Author, Journal \textbf{Volume}, page numbers (year)
%Book Author, \textit{Book title} (Publisher, place, year) page numbers
% etc
%\end{thebibliography}

\end{document}